\documentclass{article}
\usepackage[final]{graphics}
\usepackage{amsfonts,amsbsy}

\def\empile#1\over#2{\mathrel{\mathop{\kern 0pt#1}\limits_{#2}}}

\newcommand{\sll}{\raise.15ex\hbox{$/$}\kern-.43em\hbox{$l$}}
\newcommand{\slepsilon}{\raise.15ex\hbox{$/$}\kern-.53em\hbox{$\epsilon$}}
\newcommand{\slvarepsilon}{\raise.15ex\hbox{$/$}\kern-.53em\hbox{$\varepsilon$}}

\newcommand{\slL}{\raise.15ex\hbox{$/$}\kern-.53em\hbox{$L$}}
\newcommand{\slP}{\raise.15ex\hbox{$/$}\kern-.53em\hbox{$P$}}
\newcommand{\slp}{\raise.1ex\hbox{$/$}\kern-.63em\hbox{$p$}}
\newcommand{\slq}{\raise.1ex\hbox{$/$}\kern-.63em\hbox{$q$}}
\newcommand{\slv}{\raise.1ex\hbox{$/$}\kern-.63em\hbox{$v$}}
\newcommand{\slR}{\raise.15ex\hbox{$/$}\kern-.53em\hbox{$R$}}
\newcommand{\slQ}{\raise.15ex\hbox{$/$}\kern-.53em\hbox{$Q$}}
\newcommand{\slK}{\raise.15ex\hbox{$/$}\kern-.53em\hbox{$K$}}
\newcommand{\slk}{\raise.15ex\hbox{$/$}\kern-.53em\hbox{$k$}}
\newcommand{\slSigma}{\raise.15ex\hbox{$/$}\kern-.53em\hbox{$\Sigma$}}
\newcommand{\slcalP}{\raise.15ex\hbox{$/$}\kern-.63em\hbox{$\cal P$}}
\newcommand{\slA}{\raise.15ex\hbox{$/$}\kern-.73em\hbox{$A$}}
\newcommand{\slbfA}{\raise.15ex\hbox{$/$}\kern-.73em\hbox{${\imb A}$}}
\newcommand{\slpartial}{\raise.15ex\hbox{$/$}\kern-.53em\hbox{$\partial$}}

\font\tenimbf=cmmib10 at 10pt
\font\sevenimbf=cmmib10 at 7pt
\font\fiveimbf=cmmib10 at 5pt
\newfam\imbf
\textfont\imbf=\tenimbf
\scriptfont\imbf=\sevenimbf
\scriptscriptfont\imbf=\fiveimbf
\def\imb{\fam\imbf\tenimbf}

\begin{document}

\title {\bf Photon production in\\ 
high energy proton-nucleus collisions }

\author{Fran\c cois Gelis$^{(1)}$ and Jamal Jalilian-Marian$^{(2)}$}
\maketitle
\begin{center}
\begin{enumerate}
\item Laboratoire de Physique Th\'eorique,\\
B\^at. 210, Universit\'e Paris XI,\\
91405 Orsay Cedex, France
\item Physics Department,\\
         Brookhaven National Laboratory,\\
         Upton, NY 11973, USA
\end{enumerate}
\end{center}

 \begin{abstract} 
We calculate the photon production cross-section in $pA$ collisions
under the assumption that the nucleus has reached the saturation regime,
while the proton can be described by the standard parton distribution
functions. We show that due to the strong classical field $O(1/g)$ of the 
nucleus, bremsstrahlung diagrams become dominant over the direct
photon diagrams. In particular, we show that $\gamma-$jet transverse
momentum spectrum and correlations are very sensitive to gluon saturation
effects in the nucleus.
 \end{abstract}
\vskip 4mm 
\centerline{\hfill BNL-NT-02/9, LPT-ORSAY-02/40}

\section{Introduction}
An outstanding question in the description of hadronic interactions is
the problem of the parton distribution functions at very small values of
the parton momentum fraction $x$ (see \cite{Muell4,McLer1} for a
pedagogical introduction). Indeed, it is well known that the solution of
the (linear) BFKL equation \cite{Lipat1,KuraeLF1,BalitL1}, extended to
very small values of $x$, leads to cross-sections that increase
asymptotically like a power of the center of mass energy, in
contradiction with bounds coming from unitarity considerations.  It is
expected that saturation effects must limit the growth of parton
distributions at small values of $x$ \cite{GriboLR1,MuellQ1,FrankS1}.
It has been argued that this description becomes inadequate since
non-linear effects should become important at small $x$ when the
corresponding occupation numbers reach a value comparable to
$1/\alpha_{_{S}}$ \cite{GriboLR1,JalilKMW1,KovchM1,KovchM2}.

A theoretical description of this phenomenon is provided by
the Colored Glass Condensate model \cite{McLerV1,McLerV2,McLerV3}. In this
model, small-$x$ gluons are described by a classical color field, due
to the fact that these modes have a large occupation number. This color
field is driven by a classical Yang-Mills equation whose source term is
provided by faster partons. As one varies the separation scale between
the soft and hard modes, one obtains a renormalization group equation
that provides a non-linear generalization of the BFKL evolution equation
\cite{AyalaJMV1,AyalaJMV2,JalilKLW1,JalilKLW2,JalilKLW3,JalilKLW4,JalilKW1,KovneM1,KovneMW3,Balit1,Kovch3,IancuLM1,IancuLM2,Muell5}.
An important parameter of the model is the saturation momentum scale,
usually denoted $Q_s$ and defined as the transverse momentum scale below
which saturation effects become important. This scale increases
with energy and with the size of the nucleus
\cite{Muell4,JalilKMW1,KovchM1}.

More recently, some tests have been proposed in order to observe
saturation effects in high energy hadronic or nuclear collisions. Among
them are the re-analysis of HERA deep inelastic results under the
hypothesis that the $\gamma^*$-p cross-section is affected by saturation
effects \cite{GolecW1,GolecW2,GolecW3}. In the field of heavy ion
collisions, it has been suggested that one could predict the
multiplicity by describing the two nuclei with classical color fields
\cite{KovneMW1,KovneMW2,Kovch4,KrasnV1,KrasnV2}. This idea lead to a
successful description of the observed multiplicities at RHIC
\cite{KharzN1,KharzL1,KharzLN1} which is however still under debate
\cite{BaierMSS1}. It has also been proposed recently that saturation
effects could be observed in the cleaner environment provided by
ultra-peripheral heavy ion collisions \cite{GelisP1,GelisP2} since the
production of heavy quark pairs is also very sensitive to the saturation
scale. 

Various studies also proposed to study saturation effects in the
context of $pA$ collisions \cite{DumitM1,DumitJ1,DumitJ2}. In this paper, we
calculate the cross-section for photon production in $pA$ collisions and
show that the result is also sensitive to saturation effects. We assume
that the nucleus is large enough so that it can be described in terms of
classical color fields. Concerning the proton, we do not make such an
assumption and describe it as in ordinary perturbative calculations, in
terms of parton distribution functions. In this framework, the leading
order process for photon production involves a quark from the proton
that scatters off the color field of the nucleus and emits a photon by
bremsstrahlung. Therefore, the elementary process we consider in this
paper is $q\to q\gamma$ in the background of a classical color field.

The paper is organized as follows. In section \ref{sec:amplitude}, we
formulate the amplitude for the bremsstrahlung of a photon by a quark
(or antiquark) in the presence of a classical color field. This
amplitude is then evaluated. The average over the configurations of the
hard sources is performed in section \ref{sec:average}. In section
\ref{sec:Xsection}, we explain how to obtain the differential
cross-section from the semi-classical amplitude derived in section
\ref{sec:amplitude}. In section \ref{sec:diffraction}, we show that the
{\sl diffractive} (i.e. with the additional constraint that the exchange
between the quark and the nucleus is color singlet and carries a zero
total transverse momentum) cross-section for this process is
vanishing. Finally, in section \ref{sec:inclusive}, we derive the
inclusive photon radiation differential cross-section and show that the
correlations between the outgoing quark and the photon are affected by
saturation effects.

\section{Bremsstrahlung amplitude}
\label{sec:amplitude}
We want to calculate the following processes in $pA$ collisions
\begin{eqnarray}
q \,(p) + A \rightarrow q \,(q) \,\gamma (k) \,X \\
q \,(p) + A \rightarrow q \,(q) \,\gamma (k) \, A 
\label{eq:process}
\end{eqnarray}
where the quark can become either a jet or a hadron.

The starting point is the amplitude:
\begin{eqnarray}
\left<q({\imb q})\gamma ({\imb k})_{\rm out}|q({\imb p})_{\rm in}\right>=
\big<0_{\rm out}\big|a_{\rm out}({\imb k})b_{\rm out}({\imb
q})b^{\dagger}_{\rm in}({\imb p})\big|0_{\rm in}\big>
\label{eq:amp}
\end{eqnarray}
which, using the LSZ formalism \cite{ItzykZ1} can be written as
\begin{eqnarray}
&&\big<0_{\rm out}\big|a_{\rm out}({\imb k})b_{\rm out}({\imb
q})b^{\dagger}_{\rm in}({\imb p})\big|0_{\rm in}\big>
=
{e \over Z_2\sqrt{Z_3}} \int d^4x\, d^4y\, d^4z\, e^{i(k\cdot x + q\cdot
z -p\cdot y)}
\nonumber \\
&&\qquad\qquad\qquad\times\overline{u}({\imb q})(i \stackrel{\rightarrow}{\slpartial}_z -m) 
\big<0_{\rm out}\big|{\rm T} \psi (z) \varepsilon\cdot J(x)
\overline{\psi}(y)\big|0_{\rm in}\big> (i
\stackrel{\leftarrow}{\slpartial}_y +m) u({\imb p})\nonumber\\
&&
\label{eq:general}
\end{eqnarray}
where $J_{\mu}(x)\equiv \overline{\psi}(x)\gamma_{\mu}\psi (x)$ is the
quark component of the electromagnetic current, $Z_2$ and $Z_3$ are the
fermion and photon wave function renormalization factors\footnote{Since
we are not including loop corrections in this calculation, we will
simply set these factors to $1$ in the following.}, and $e$ is the
fractional electric charge of the quark under consideration. So what we
basically need is
\begin{eqnarray}
\big<0_{\rm out}\big|{\rm T} \psi (z) \overline{\psi}(x) {\slvarepsilon}
\psi (x) \overline{\psi}(y)\big|0_{\rm in}\big>
\end{eqnarray}
which, using Wick's theorem, can be written as
\begin{eqnarray}
\big<0_{\rm out}\big|{\rm T} \psi (z) \overline{\psi}(x) {\slvarepsilon}
\psi (x) \overline{\psi}(y)\big|0_{\rm in}\big> = - G_{_{F}}(z,x) {\slvarepsilon} G_{_{F}}(x,y)
\end{eqnarray}
where the fermion Feynman propagator $G_{_{F}}$ is defined
as\footnote{Compared to the definition of the fermion propagator in
\cite{GelisP1,BaltzGMP1}, we are omitting the denominator $\left<0_{\rm
out}|0_{\rm in}\right>$ due to the fact that this factor is a pure phase
in the case where the background contains the classical field of a
single nucleus. This phase indeed drops out of physical quantities
obtained after squaring amplitudes.}
\begin{eqnarray}
G_{_{F}}(x,y)\equiv \big<0_{\rm out}\big|{\rm T} \overline{\psi} (y)
{\psi}(x) \big|0_{\rm in}\big>\; .
\end{eqnarray}
It should be noted that $G_{_{F}}$ is the fermion propagator in the 
background of the strong classical color field of the nucleus.
Our amplitude then becomes
\begin{eqnarray}
&&
\big<q({\imb q})\gamma ({\imb k})_{\rm out}\big|q({\imb p})_{\rm in}\big>=
e \int d^4x\, d^4y\, d^4z\, e^{i(k\cdot x + q\cdot z -p\cdot
y)}\nonumber\\
&&\qquad\qquad\times
\overline{u}({\imb q})(i \stackrel{\rightarrow}{\slpartial}_z -m) 
G_{_{F}}(z,x) {\slvarepsilon} G_{_{F}}(x,y)
(i \stackrel{\leftarrow}{\slpartial}_y -m) u({\imb p})\; .
\label{eq:amplitude}
\end{eqnarray}

The fermion propagator $G_{_{F}}$ is already known from
\cite{GelisP1,GelisP2,McLerV4,HebecW1}.  In coordinate space and in the
``singular'' gauge, it is given by the following formula \cite{McLerV4}
\begin{eqnarray}
G_{_{F}}(x,y) &=& G_{_{F}}^{0}(x-y) + \int d^4z \delta(z^-) 
\bigg[\theta (x^-) \theta (-y^-)(U^{\dagger}({\imb z}_\perp) -1)\nonumber \\
&& - \theta (-x^-)\theta (y^-)(U({\imb z}_\perp) -1)\bigg]
G_{_{F}}^{0}(x-z) \gamma^- G_{_{F}}^{0}(z-y)\; ,
\label{eq:G_Fcord}
\end{eqnarray}
where $G_{_{F}}^0$ is the free Feynman propagator of a quark, and where
$U({\imb z}_\perp)$ is a unitary matrix containing the interactions
between the quark and the colored glass condensate.  It is convenient to
separate out the interaction part of the propagator and write it in
momentum space as
\begin{eqnarray}
G_{_{F}}(q,p)= (2\pi)^4 \delta^4 (q-p) G_{_{F}}^0 (p) + G_{_{F}}^0 (q)
{\cal T}_{_{F}} (q,p) G_{_{F}}^0 (p)
\label{eq:Gint}
\end{eqnarray}
where 
\begin{eqnarray}
{\cal T}_{_{F}} (q,p)=2\pi \delta (q^- - p^-) \gamma^- 
{\rm sign} (p^-) \!\!\int\! d^2 {\imb z}_\perp \bigg [ U^{{\rm sign} (p^-)}
({\imb z}_\perp) -1 \bigg ]
e^{i({\imb q}_\perp - {\imb p}_\perp)\cdot {\imb z}_\perp}
\label{eq:taures}
\end{eqnarray}
while for a nucleus moving in the positive $z$ direction we have
\begin{eqnarray}
U({\imb x}_\perp) \equiv {\rm T} \exp \bigg \{-ig^2 \int^{+\infty}_{-\infty}
d z^- {1 \over {\nabla^2_\perp}} \rho_a (z^-,{\imb z}_\perp) t^a \bigg\}
\label{eq:Udef}
\end{eqnarray}
with $t^a$ in the fundamental representation, and where
$\rho_a(z^-,{\imb z}_\perp)$ is the density of color sources in the nucleus. 

Inserting this propagator in the expression for the photon production
amplitude (\ref{eq:amplitude}) , we find the following expression in
momentum space 
\begin{eqnarray}
&&
\big<q({\imb q})\gamma ({\imb k})_{\rm out}\big|q({\imb p})_{\rm in}\big>=
-e\, \overline{u}({\imb q}) \,\Bigg[ (2\pi)^4 \delta^4 (k+q-p)    
{\slvarepsilon} \nonumber\\
&&\qquad\qquad+
{\cal T}_{_{F}}(q,p-k) \,G_{_{F}}^{0}(p-k)\,
{\slvarepsilon}
 + {\slvarepsilon}\,
G_{_{F}}^{0}(q+k)\,{\cal T}_{_{F}}(q+k,p)\nonumber \\
&&
\qquad\qquad+
\int {{d^4l} \over {(2\pi)^4}} {\cal T}_{_{F}}(q,l)\,G_{_{F}}^{0}(l) 
\,{\slvarepsilon}\,G_{_{F}}^{0}(k+l)\,{\cal T}_{_{F}}(k+l,p)
\Bigg]\,u({\imb p})\; .
\nonumber\\
&&
\label{eq:M} 
\end{eqnarray}
The first term in (\ref{eq:M}) corresponds to emission of a photon by
the quark with no scattering from the Color Glass Condensate. This term
cannot contribute to the cross section for real photons and indeed
vanishes (the delta function $\delta(p-q-k)$ has no support for on-shell
particles).

\begin{figure}[ht]
\centerline{\resizebox*{!}{1.5cm}{\includegraphics{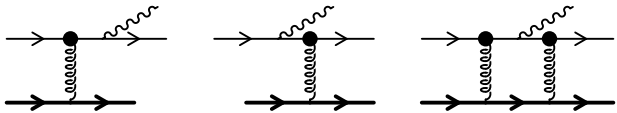}}}
\caption{\label{fig:processes}The physical processes contributing to the
bremsstrahlung of a quark in a colored glass condensate. The black dot
denotes the interaction of a quark to all orders with the background
field, i.e. the scattering matrix ${\cal T}_{_{F}}$ defined in
Eq.~(\ref{eq:taures}).}
\end{figure}

The second and third term,respectively, correspond to the case when the
quark multiply scatters from the Color Glass Condensate before and after
emission of the photon. The last term describes the case when the quark
multiply scatters from the Color Glass Condensate, then emits a photon
and again, multiply scatters from the Color Glass Condensate. Those
processes are illustrated in the figure \ref{fig:processes}. Note that
Eq.~(\ref{eq:M}) is an exact formula for the bremsstrahlung of a quark
in a colored glass condensate. Indeed, it resums the interactions to
all orders with the classical background field, as required by the large
gluon density in the nucleus.

Now we will show that the last term also vanishes. To show this, it is
sufficient to consider the structure of the poles in the propagators in
the last term. One has the following structure
\begin{eqnarray}
\delta (q^- - l^-) \delta (k^- + l^- - p^-) \!\!\int\limits_{-\infty}^{+\infty}\!\! dl^+\!\!
{1 \over l^+ - {{\vphantom{({\imb l}_\perp+{\imb k}_\perp)^2}}{\imb l}_\perp^2 + m^2 -i\epsilon \over 2l^-}}\,
{1 \over l^+ + k^+ - {({\imb l}_\perp+{\imb k}_\perp)^2 +m^2 -i\epsilon \over 2(k^-+l^-)}}
\label{eq:poles}
\end{eqnarray}
Since both $p^- > 0$ and $q^- > 0$, then $l^- > 0$ as well as $k^-+l^- >
0$. This means that both $l^+$ poles are below the real axis. Therefore,
one can close the integration contour above the real axis and get a
vanishing contribution. The only exception to this argument would occur
if there were terms
proportional to $l^+$ in the numerator which can compensate the $l^+$ in
the denominator, a situation in which the theorem of residues would give
a nonzero contribution from the half circle at infinity used to close
the contour. To investigate this case, we write explicitly the terms which
contain $l^+$ in the numerator of the last term. We have
\begin{eqnarray}
\overline{u}({\imb q}) \gamma^- \;{\sll}\;\; {\slvarepsilon}\;\;
{\sll}\; \gamma^- u({\imb p})
= l^+ l^+ \overline{u}({\imb q}) \gamma^- \gamma^- \;\; {\slvarepsilon}\;\; 
\gamma^- \gamma^-  u({\imb p})\; .
\label{eq:num}
\end{eqnarray}
Since $\gamma^- \gamma^- =0$, equation (\ref{eq:num}) identically vanishes.
Therefore, the last term in the amplitude (\ref{eq:M}) is zero. That
this term vanishes has a simple physical explanation. Indeed, in this
term the quark first scatters off the nucleus, then propagates for a
while and emits the photon, then propagates again and finally
reinteracts with the nucleus. Note that the intermediate free
propagations are off-shell and take some nonzero amount of time. But
since the nucleus moves at the speed of light in the $+z$ direction in
our model, it is already far away behind the quark by the time the
photon has been emitted, and a second scattering of the quark off the
nucleus cannot happen\footnote{One can notice that the argument used to
prove that Eq.~(\ref{eq:poles}) vanishes would not be applicable if the
nucleus was moving at a speed $v<1$. Indeed, in that case we would not
have the delta functions implying the conservation of the $-$ component
of the quark momentum, and we could have poles on both sides of the real
$l^+$ axis.}.

The amplitude can then be written as
\begin{eqnarray}
&&\big<q({\imb q})\gamma ({\imb k})_{\rm out}\big|q({\imb p})_{\rm
in}\big>=\nonumber\\
&&\qquad= -e\, \overline{u}({\imb q})\,\Bigg[ {\cal T}_{_{F}}(q,p-k)
\,G_{_{F}}^{0}(p-k)\, {\slvarepsilon} +{\slvarepsilon}\, G_{_{F}}^{0}(q+k)\,{\cal T}_{_{F}}(q+k,p) \Bigg]\,
u({\imb p})\nonumber\\
&&\qquad= -ie\,\overline{u}({\imb q})\bigg[ 
{\gamma^- ({\slp} - {\slk} + m) {\slvarepsilon} \over (p-k)^2 -m^2} +
{{\slvarepsilon}\;({\slq} + {\slk} + m)\gamma^- \over (q+k)^2 -m^2}
\bigg]u({\imb p})\nonumber\\
&&\qquad\qquad\times2\pi\delta(q^-+k^--p^-)\int d^2{\imb x}_\perp
e^{i({\imb q}_\perp+{\imb k}_\perp-{\imb p}_\perp)\cdot{\imb x}_\perp}
\Big(U({\imb x}_\perp)-1\Big)\; .
\label{eq:finalamp}
\end{eqnarray}
Depending on whether one wants to calculate a diffractive
or inclusive quantity, one would need to color average and then
square the amplitude or square the amplitude and then color
average. In either case, the color averaging does not affect
the spinor structure and one can evaluate the spinor dependence
and color averaging independently.

The spinor dependence of the amplitude (\ref{eq:finalamp}) squared 
is given by
\begin{eqnarray}
\left<{\rm tr}(L^{\dagger} L)\right>_{\rm spin} &\equiv &{1\over 2}{\rm tr} \Bigg\{ ({\slp} + m) \bigg[
{{\slvarepsilon}^{\ast}\;({\slp} - {\slk} + m)\gamma^- 
\over  (p-k)^2 - m^2 } + 
{\gamma^- ({\slq} + {\slk} + m)\;{\slvarepsilon}^{\ast} \over (q+k)^2 -m^2}
\bigg] \nonumber \\
&&\quad\times ({\slq} + m) \bigg[
{\gamma^- ({\slp} - {\slk} + m) {\slvarepsilon} \over (p-k)^2 -m^2} +
{{\slvarepsilon}\;({\slq} + {\slk} + m)\gamma^- \over (q+k)^2 -m^2}
\bigg] \Bigg\}
\label{eq:spinsq} 
\end{eqnarray}
where the factor $1/2$ comes from averaging of the spin of the incoming
quark.  Summing also over the spin of the final quark and over the
polarization of the photon, a straightforward calculation gives:
\begin{eqnarray}
\left<{\rm tr}(L^{\dagger} L)\right>_{\rm spin}&=&
-4m^2\left[
{{p^-{}^2}\over{(q\cdot k)^2}}
+{{q^-{}^2}\over{(p\cdot k)^2}}
+{{k^-{}^2}\over{(p\cdot k)(q\cdot k)}}
\right]\nonumber\\
&&+8(p^-{}^2+q^-{}^2)\left[
{{p\cdot q}\over{(p\cdot k)(q\cdot k)}}
+{1\over{q\cdot k}}-{1\over{p\cdot k}}
\right]\; .
\end{eqnarray}

\section{Color averages}
\label{sec:average}
In order to perform the color averaging of this amplitude with the weight
\begin{equation}
W[\rho]\equiv\exp\Bigg\{ -\int dz^- d^2{\imb z}_\perp {{\rho_a(z^-,{\imb
z}_\perp)\rho^a(z^-,{\imb z}_\perp)}\over{2\mu^2(z^-)}}\Bigg\}
\end{equation}
 we will need to evaluate
expressions like $\left<U({\imb x}_\perp)\right>_{\rho}$ and
$\left<U^{\dagger}({\imb x}_\perp)\,U({\imb y}_\perp)\right>_{\rho}$
where $U({\imb x}_\perp)$ is given by (\ref{eq:Udef}). These are
already evaluated in \cite{GelisP1,GelisP2}. Following the notations
of these papers, we obtain 
\begin{eqnarray}
\left< U({\imb x}_\perp)-1 \right>_{\rho} = {\cal P}({\imb
x}_\perp)(e^{-B_1}-1)
\label{eq:Uave}
\end{eqnarray}
where ${\cal P}({\imb x}_\perp)$ is a function that describes the
transverse profile of the nucleus. It can be thought of as a function
whose value is $0$ outside the nucleus and $1$ inside the nucleus. The
object $B_1$ appearing in this expression is given by
\begin{equation}
B_1({\imb x}_\perp)\equiv Q_s^2 \int d^2{\imb z}_\perp G_0^2({\imb x}_\perp-{\imb z}_\perp)\sim {{Q_s^2}\over{\Lambda_{_{QCD}}^2}}\; ,
\end{equation}
with $Q_s^2\equiv g^4 (t_at^a)\int_{-\infty}^{+\infty}dz^- \mu^2(z^-)/2$
the saturation scale\footnote{ The saturation momentum would acquire a 
dependence on the rapidity of the quark via quantum evolution effects not 
included explicitly here. Indeed, the quark is sensitive to all of the
nucleus constituents that have a rapidity between the quark
rapidity and the nucleus rapidity.} (the integral of $\mu^2$ over $z^-$
is the number density of color sources per unit of transverse area in
the target nucleus).  Similarly, we have
\begin{eqnarray}
\left< (U^{\dagger}({\imb x}_\perp)-1)( U({\imb y}_\perp)-1) \right>_{\rho} =
{\cal P}({\imb x}_\perp){\cal P}({\imb y}_\perp)\Big[
1+e^{-B_2({\imb x}_\perp-{\imb y}_\perp)}-2e^{-B_1}
\Big]
\label{eq:UUave}
\end{eqnarray}
with the definition
\begin{eqnarray}
B_2({\imb x}_\perp-{\imb y}_\perp)&\equiv&
Q_s^2\int d^2{\imb z}_\perp [G_0({\imb x}_\perp-{\imb z}_\perp)
-G_0({\imb y}_\perp-{\imb z}_\perp)]^2\nonumber\\
&\approx& {{Q_s^2({\imb x}_\perp-{\imb y}_\perp)^2}\over{4\pi}}\ln\Big( {1\over{|{\imb x}_\perp-{\imb y}_\perp|\Lambda_{_{QCD}}}}\Big)\; .
\label{eq:B2}
\end{eqnarray}
In the above equations, $G_0 ({\imb z}_\perp - {\imb y}_\perp)$ is the
free propagator in two dimensions, defined by
\begin{eqnarray}
{\partial^2 \over \partial {\imb z}_\perp^2} G_0 ({\imb z}_\perp - {\imb
y}_\perp) = \delta({\imb z}_\perp - {\imb y}_\perp)
\end{eqnarray}
and given explicitly by
\begin{eqnarray}
G_0 ({\imb z}_\perp - {\imb y}_\perp) = - \int {d^2 {\imb k}_\perp \over (2\pi)^2}
{e^{i{\imb k}_\perp\cdot({\imb z}_\perp - {\imb y}_\perp)} \over {\imb
k}_\perp^2}\; .
\label{eq:G0}
\end{eqnarray} 
Note that the objects evaluated in Eqs.~(\ref{eq:Uave}) and
(\ref{eq:UUave}) are matrices in the fundamental representation of
$SU(N_c)$ that are proportional to the unit matrix. In the calculation
of cross-sections, one must sum over the color of the outgoing quark and
average over the color of the incoming quark, which amounts to taking the
color trace of this matrix and dividing by $N_c$. Therefore, 
Eqs.~(\ref{eq:Uave}) and (\ref{eq:UUave}) can be seen as scalars giving
directly the result of this procedure.

\section{Cross-Section}
\label{sec:Xsection}
At first sight, the square of the delta function $\delta (p^- - k^- -
q^-)$ that appears when we square the amplitude might seem a little
worrisome. However, this is just a manifestation of Fermi's Golden
rule, adapted to the symmetries of the present problem. Indeed, the
target nucleus being invariant under (light-cone) time $x^+$, the $-$
component of the projectile momentum is conserved, hence the $\delta
(p^- - k^- - q^-)$ at the amplitude level. Then, the $2\pi\delta(0^-)$
we have in the amplitude squared is just an artifact of shooting a
plane wave at this target instead of a properly normalized wave
packet. It should be interpreted as a large but finite time $\Delta
x^+$ (see \cite{ItzykZ1}, pages $96-97$ and \cite{BjorkD1}, pages
$100-102$ for a nice discussion of this).  It is in fact simpler to
follow \cite{PeskiS1} (pages 99-107) and introduce a wave packet
instead of the incoming plane wave
\begin{equation}
\big|\phi_{\rm in}\big>\equiv \int {{d^3{\imb l}}\over{(2\pi)^3}} {{e^{i{\imb
b}\cdot{\imb l}_\perp}}\over{\sqrt{2E_{\imb l}}}} \phi({\imb l})\big|q({\imb
l})_{\rm in}\big>\; ,
\end{equation}
where $\phi({\imb l})$ is some wave packet that is peaked around the
central value ${\imb p}$. The phase $\exp(i{\imb b}\cdot{\imb
l}_\perp)$ is here to account for a finite impact parameter between
the wave packet of the quark and the trajectory of the center of the
nucleus. Normalization is chosen in such a way that:
\begin{equation}
\big<\phi_{\rm in}\big|\phi_{\rm in}\big>=1\; ,\qquad{\rm i.e.}\; 
\int{{d^3{\imb l}}\over{(2\pi)^3}} \big|\phi({\imb l})\big|^2=1\; .
\end{equation}
Then, the differential (per unit of volume in the invariant phase
space of the final state particles) interaction probability between
this wave packet and the nucleus is simply
\begin{equation}
dP(b)\equiv{{d^3{\imb k}}\over{(2\pi)^32k_0}}{{d^3{\imb q}}\over{(2\pi)^32q_0}}
\left|\big<q({\imb q})\gamma({\imb k})_{\rm out}\big|
\phi_{\rm in}\big>\right|^2\; .
\end{equation}
Recalling now the relation between the differential cross-section and
the probability defined in the previous equation
\begin{equation}
d\sigma=\int d^2{\imb b}\; dP({\imb b})\; ,
\end{equation}
we have for the cross-section:
\begin{eqnarray}
&&d\sigma={{d^3{\imb k}}\over{(2\pi)^32k_0}}{{d^3{\imb q}}\over{(2\pi)^32q_0}}
\int d^2{\imb b} {{d^3{\imb l}}\over{(2\pi)^3}} 
{{d^3{\imb l^\prime}}\over{(2\pi)^3}}
e^{i{\imb b}\cdot({\imb l}_\perp-{\imb l^\prime}_\perp)}
{{\phi({\imb l})}\over{\sqrt{2E_{\imb l}}}}
{{\phi({\imb l^\prime})^*}\over{\sqrt{2E_{\imb l^\prime}}}}\nonumber\\
&&\qquad\qquad\qquad\qquad\times
\big<q({\imb q})\gamma({\imb k})_{\rm out}\big|
q({\imb l})_{\rm in}\big>
\big<q({\imb l^\prime})_{\rm in}\big|q({\imb q})\gamma({\imb k})_{\rm out}\big>\; .
\end{eqnarray}
Performing the integration over the impact parameter ${\imb b}$ brings
a $(2\pi)^2\delta({\imb l}_\perp-{\imb l^\prime}_\perp)$.  If we
factor out the delta function contained in the amplitude in the
following way
\begin{equation}
\big<q({\imb q})\gamma({\imb k})_{\rm out}\big|
q({\imb l})_{\rm in}\big>\equiv 2\pi\delta(l^--q^--k^-) {\cal M}({\imb l}|{\imb q}{\imb k})\; ,
\end{equation}
we can also perform the integral over $d^3{\imb l^\prime}$ by using
one of those delta functions and get a factor $\sqrt{2E_{\imb
p}}/2p^{-}$, where we have taken advantage of the fact that the wave
packets are peaked around ${\imb p}$ to replace by ${\imb p}$ the
dummy variables ${\imb l}$ and ${\imb l^\prime}$ whenever they are in
slowly varying quantities (i.e. everywhere except in the wave-packets
themselves). At this stage, the differential cross-section reads
\begin{eqnarray}
d\sigma={{d^3{\imb k}}\over{(2\pi)^32k_0}}{{d^3{\imb q}}\over{(2\pi)^32q_0}}
{1\over{2p^-}} \left|{\cal M}({\imb p}|{\imb q}{\imb k})\right|^2 2\pi\delta(p^--q^--k^-)\; .
\label{eq:generic-cs}
\end{eqnarray}
This is the formula we are going to use in the rest of this paper.

\section{Diffraction}
\label{sec:diffraction}
In order to calculate the diffractive cross section, we average the
amplitude (\ref{eq:finalamp}) over color charge and then square
it. This involves evaluating expressions like $\left<U({\imb x}_\perp) -
1\right>_{\rho}$ which is given by Eq.~(\ref{eq:Uave}). Performing also the
integrals with respect to the transverse positions, we obtain
\begin{eqnarray}
\left|{\cal M}({\imb p}|{\imb q}{\imb k})\right|^2_{\rm diff} \!&=&\!
e^2 \left<{\rm tr}(L^{\dagger} L)\right>_{\rm spin}
\bigg[1-e^{-B_1}\bigg]^2 \nonumber\\
&&\quad\times
\widetilde{\cal P}({\imb p}_\perp-{\imb q}_\perp-{\imb k}_\perp)
\widetilde{\cal P}(-{\imb p}_\perp+{\imb q}_\perp+{\imb k}_\perp)\; ,
\label{eq:diffampsq}
\end{eqnarray}
where the function $\widetilde{\imb P}({\imb l}_\perp)$ is the Fourier
transform of the profile function ${\cal P}({\imb x}_\perp)$. Note that
for a large nucleus, this Fourier transform is very sharply peaked
around ${\imb l}_\perp=0$, with a typical width of $1/R$ where $R$ is
the radius of the nucleus. Since this momentum scale is much smaller
than any other typical momentum scale in the problem, we can
approximate
\begin{eqnarray}
\widetilde{\cal P}({\imb p}_\perp-{\imb q}_\perp-{\imb k}_\perp)
\widetilde{\cal P}(-{\imb p}_\perp+{\imb q}_\perp+{\imb k}_\perp)
&\approx& \widetilde{\cal P}(0)(2\pi)^2\delta({\imb p}_\perp-{\imb q}_\perp-{\imb k}_\perp)\nonumber\\
&=&\pi R^2(2\pi)^2\delta({\imb p}_\perp-{\imb q}_\perp-{\imb k}_\perp)\; .\nonumber\\
&&
\label{eq:PP}
\end{eqnarray}
Therefore, we can rewrite the square of the diffractive amplitude as
\begin{eqnarray}
\left|{\cal M}({\imb p}|{\imb q}{\imb k})\right|^2_{\rm diff}\!=\!
e^2 \pi R^2 \left<{\rm tr}(L^{\dagger} L)\right>_{\rm spin}\!\bigg[1-e^{-B_1}\bigg]^2\!
(2\pi)^2\delta({\imb p}_\perp-{\imb q}_\perp-{\imb k}_\perp)\; ,
\end{eqnarray}
so that the diffractive cross-section for bremsstrahlung becomes:
\begin{eqnarray}
&&d\sigma^{q\to q\gamma}_{\rm diff}={{d^3{\imb
k}}\over{(2\pi)^32k_0}}{{d^3{\imb q}}\over{(2\pi)^32q_0}}
{{e^2 \pi R^2}\over{2p^-}}
\left<{\rm tr}(L^{\dagger} L)\right>_{\rm spin}
\left[1-e^{-B_1}\right]^2
\nonumber\\
&&\qquad\qquad\qquad\times
(2\pi)^3 \delta({\imb p}_\perp-{\imb q}_\perp-{\imb k}_\perp)
\delta(p^--q^--k^-)\; .
\label{eq:diff-cs}
\end{eqnarray}
To proceed further, we assume that the incoming quark is massless
and that its transverse momentum ${\imb p}_\perp$ is zero. 
This is reasonable if we have in mind an application of this
calculation to photon production in $pA$ collisions. Indeed, there is no
significant amount of heavy quarks in the wave function of the proton
and partons are collinear to the proton. Under these assumptions, it is
a matter of straightforward algebra to show that 
\begin{eqnarray}
\left<{\rm tr}(L^{\dagger} L)\right>^{\rm diff}_{\rm spin} = 0
\end{eqnarray}
leading to
\begin{eqnarray}
{{d\sigma^{q\to q\gamma}_{\rm diff}}\over{d^2{\imb k}_\perp}}=0.
\end{eqnarray}
This result is in fact rather intuitive: asking for a diffractive
process implies that no net transverse momentum is exchanged between the
classical color field and the quark, which inhibits the radiation of a
photon by bremsstrahlung.

\section{Inclusive}
\label{sec:inclusive}
In order to calculate the inclusive cross section, we square the amplitude
and then average over color charge. 
Using (\ref{eq:UUave}), we obtain for the inclusive amplitude squared
\begin{eqnarray}
\left|{\cal M}({\imb p}|{\imb q}{\imb k})\right|^2_{\rm incl}&=&  e^2
\left<{\rm tr}(L^{\dagger} L)\right>_{\rm spin}
\int d^2{\imb x}_\perp\, d^2{\imb y}_\perp\,e^{i({\imb q}_\perp+{\imb
k}_\perp-{\imb p}_\perp)\cdot({\imb x}_\perp-{\imb y}_\perp)}\nonumber\\
&&\qquad\times
{\cal P}({\imb x}_\perp){\cal P}({\imb y}_\perp)\left[
1+e^{-B_2({\imb x}_\perp-{\imb y}_\perp)}-2e^{-B1}
\right]\; .
\end{eqnarray}
By inspecting the formula (\ref{eq:B2}), we see that $\exp(-B_2)$
becomes very small if $|{\imb x}_\perp-{\imb y}_\perp|$ becomes larger
than $1/Q_s$. Therefore, this term contributes only if the separation
between ${\imb x}_\perp$ and ${\imb y}_\perp$ is much smaller than the
radius of the nucleus (assuming a large nucleus such that $R\gg
Q_s^{-1}$). Therefore, for this term we can approximate
\begin{equation}
{\cal P}({\imb x}_\perp){\cal P}({\imb y}_\perp)\approx
{\cal P}^2({\imb x}_\perp)\approx {\cal P}({\imb x}_\perp)\; ,
\end{equation}
  where the last equality is valid for a nucleus with rather sharp
  edges (i.e. if ${\cal P}({\imb x}_\perp)$ goes from $0$ to $1$ rather
  fast). Using Eq.~(\ref{eq:PP}) and performing the Fourier transforms,
  we can rewrite the inclusive amplitude squared as follows
\begin{eqnarray}
&&
\left|{\cal M}({\imb p}|{\imb q}{\imb k})\right|^2_{\rm incl}=  e^2
\pi R^2
\left<{\rm tr}(L^{\dagger} L)\right>_{\rm spin}\nonumber\\
&&\qquad\qquad\quad\times
\left[D({\imb p}_\perp-{\imb q}_\perp-{\imb
k}_\perp)+ (1- 2 e^{-B_1})(2\pi)^2\delta({\imb p}_\perp-{\imb
q}_\perp-{\imb k}_\perp)\right]\; ,\nonumber\\
&&
\end{eqnarray}
where, following \cite{GelisP2}, we denote
\begin{equation}
D({\imb l}_\perp)\equiv \int d^2{\imb x}_\perp e^{i{\imb
l}_\perp\cdot{\imb x}_\perp} (e^{-B_2({\imb x}_\perp)}-1)=
\int d^2{\imb x}_\perp e^{i{\imb
l}_\perp\cdot{\imb x}_\perp}
\left<U(0)U^\dagger({\imb x}_\perp)-1\right>_\rho\; .
\end{equation}
Inserting this in the formula (\ref{eq:generic-cs}), we obtain the
following formula for the inclusive cross-section\footnote{The invariant 
phase space of a massless
particle can be rewritten in terms of the light-cone variables as follows:
\begin{equation}
{{d^3{\imb k}}\over{(2\pi)^3 2k_0}}={{dk^+ dk^- d^2{\imb
k}_\perp}\over{(2\pi)^4}}
2\pi\theta(k^{\pm})\delta(2k^- k^+ -{\imb k}_\perp^2)\; .
\end{equation}} 
\begin{eqnarray}
&&d\sigma^{q\to q\gamma}_{\rm incl}={{d^3{\imb
k}}\over{(2\pi)^32k_0}}{{d^3{\imb q}}\over{(2\pi)^32q_0}}
{{e^2 \pi R^2 }\over{2p^-}}
\left<{\rm tr}(L^{\dagger} L)\right>_{\rm spin}2\pi\delta(p^--q^--k^-)
\nonumber\\
&&\qquad\qquad\times 
\left[D({\imb p}_\perp-{\imb q}_\perp-{\imb
k}_\perp)+ (1- 2 e^{-B_1})(2\pi)^2\delta({\imb p}_\perp-{\imb
q}_\perp-{\imb k}_\perp)\right]\; .\nonumber\\
&&
\label{eq:inclusive-cs}
\end{eqnarray}
Another simplification can be achieved by neglecting the term in
$\exp(-B_1)$ since  $B_1 \sim
Q_s^2/\Lambda_{_{QCD}}^2 \gg 1$ appears in the exponential with a
negative sign. If one introduces \cite{GelisP1} 
\begin{equation}
C({\imb l}_\perp)\equiv \int d^2{\imb x}_\perp e^{i{\imb
l}_\perp\cdot{\imb x}_\perp} e^{-B_2({\imb x}_\perp)}=
\int d^2{\imb x}_\perp e^{i{\imb
l}_\perp\cdot{\imb x}_\perp}
\left<U(0)U^\dagger({\imb x}_\perp)\right>_\rho\; ,
\label{eq:C-def}
\end{equation}
the inclusive cross section can be rewritten as
\begin{eqnarray}
&&d\sigma^{q\to q\gamma}_{\rm incl}={{d^3{\imb
k}}\over{(2\pi)^32k_0}}{{d^3{\imb q}}\over{(2\pi)^32q_0}}
{{e^2 \pi R^2}\over{2p^-}}
\left<{\rm tr}(L^{\dagger} L)\right>_{\rm spin}
\nonumber\\
&&\qquad\qquad\times 
2\pi\delta(p^--q^--k^-)\,C({\imb p}_\perp-{\imb q}_\perp-{\imb
k}_\perp)\; .\nonumber\\
&&
\label{eq:inclusive-cs1}
\end{eqnarray}
Assuming again that the incoming quark transverse momentum ${\imb
p}_\perp$ is zero and neglecting the quark mass, one can perform 
the integrals over $q^+, k^+, q^-$ using the delta
functions. There is however a complication due to collinear
singularities, i.e. singularities that show up when the emitted photon
is parallel to the outgoing quark. It is convenient to trade the
transverse momentum of the final quark for the total transverse momentum
of the final state, i.e. ${\imb l}_\perp\equiv {\imb q}_\perp+{\imb
k}_\perp$. In terms of this new variable, we have
\begin{eqnarray}
&&{1\over{\pi R^2}}
{d\sigma^{q\to q\gamma}_{\rm incl} \over d^2{\imb k}_\perp} = 
{{e^2} \over {(2\pi)^5} {\imb k}_\perp^2}
\int_0^1 dz {[1 + (1-z)^2] \over z} 
\int d^2{\imb l}_\perp {{{\imb l}_\perp^2 \,C({\imb l}_\perp )} 
\over [{\imb l}_\perp - {\imb k}_\perp/z]^2}
\label{eq:inccs} 
\end{eqnarray}
where $z\equiv k^-/p^-$ and $ {[1 + (1-z)^2]/z}$ is the standard leading
order photon splitting function. Eq. (\ref{eq:inccs}) is our main result. 
Note that $C({\imb l}_\perp)$ behaves
like $1/{\imb l}_\perp^4$ at large ${\imb l}_\perp$ which ensures that
the integral converges at large momentum transfer. In this formula,
$C({\imb l}_\perp)$ is the only object that depends on the structure of
the color sources describing the target nucleus. In particular, all the
quantum evolution effects would go into this object via the averaging
procedure in Eq.~(\ref{eq:C-def}).  One can also note that this result
exhibits the standard collinear denominator $[{\imb l}_\perp - {\imb
k}_\perp/z]^2$ that vanishes if the photon is emitted collinearly to the
quark. This aspect of the result is of course not affected by the
description of the target nucleus as a color glass condensate.

In the soft photon limit, one can see the decoupling of the photon
emission subprocess from the quark scattering part. The latter agrees
with the quark-nucleus scattering cross-section calculated in \cite{DumitJ1}.

It is instructive to perform the ``perturbative limit'' of this
result. This regime is reached when the transverse momentum ${\imb
l}_\perp$ transferred between the nucleus and the quark is large compared 
to the saturation momentum $Q_s$. In this limit, we have \cite{GelisP1}
\begin{equation}
C({\imb l}_\perp)\approx {{2Q_s^2}\over{{\imb l}_\perp^4}}\; .
\end{equation}
Using this result, we have:
\begin{equation}
{d\sigma^{q\to q\gamma}_{\rm incl} \over d^2{\imb k}_\perp}\Big|_{\rm
pert.}=
{{2N_h e_q^2 \alpha_{\rm em}
\alpha_{_{S}}^2}\over{\pi^2}}{{C_{_{F}}}\over{{\imb k}_\perp^2}}
\int\limits_0^1 dz {{1+(1-z)^2}\over{z}} \int {{d^2{\imb
l}_\perp}\over{{\imb l}_\perp^2[{\imb l}_\perp-{\imb k}_\perp/z]^2}}
\end{equation}
where $e_q$ is the quark electric charge in units of the electron
charge, and where $N_h\equiv \pi R^2\int dz^- \mu^2(z^-)$ is the total
number of hard color sources in the target nucleus. Therefore, this
expression has all the features of the bremsstrahlung of a photon 
by a quark scattering off a parton inside the nucleus with the exchange
of a gluon in the t-channel (this term is the dominant one at large
center of mass energy).  

\begin{figure}
\centerline{\resizebox*{9.5cm}{!}{\includegraphics{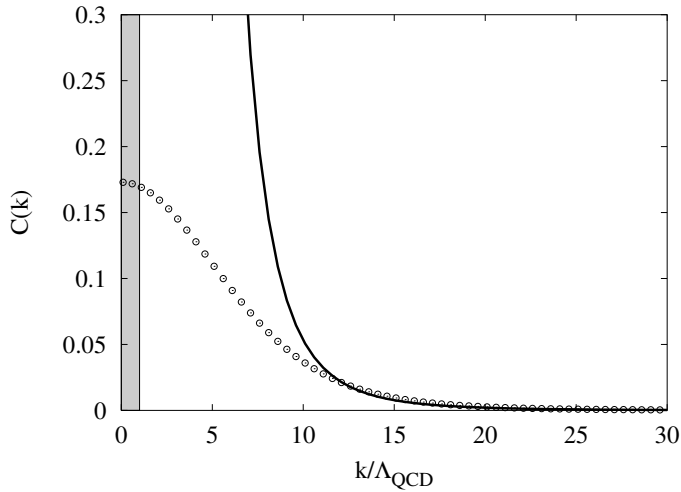}}}
\caption{\label{fig:Ck} Behavior of the correlator $C({\imb k}_\perp)$ 
  as a function of ${\imb k}_\perp$. In this plot, the value of $Q_s$
  is such that $Q_s/\Lambda_{_{QCD}}=10$. Circles: computed value of
  $C({\imb k}_\perp)$. Solid line: the ``perturbative'' value $C({\imb
    k}_\perp)\approx 2Q_s/k_\perp^4$, valid if $k_\perp\gg Q_s$.}
\end{figure}

In Eq.~(\ref{eq:inccs}), the only factor that depends crucially on the
saturation hypothesis for the nucleus is the factor $C({\imb
  l}_\perp)$. Indeed, this term contains all the dependence on the
saturation scale, as well as the modifications of the transverse
momentum spectrum at scales below $Q_s$. The transverse momentum
dependence of this object is illustrated in figure \ref{fig:Ck}.
In order to observe effects due to this factor, it would be useful to
measure both the radiated photon and the jet induced by the outgoing
quark. The photon-jet correlations, and in particular the distribution
of their total transverse momentum ${\imb l}_\perp={\imb
  q}_\perp+{\imb k}_\perp$, would indeed enable one to extract in a
rather direct way the function $C({\imb l}_\perp)$ itself. On the
contrary, if one measures only the photon spectrum, one can access
only a given moment of this function.

\section{Conclusions and perspectives}
In this paper, we have studied the photon production in $pA$ collisions
by bremss\-trah\-lung of a quark in the saturated color field of the
nucleus. We have shown that the cross-section for this process depends
sensitively on saturation effects. Furthermore, we have shown that
photon bremsstrahlung diagrams which are, in the standard perturbation
theory, higher order compared to the direct photon diagrams, become
leading order. This is due to the presence of a strong color field
$O(1/g)$ in the target nucleus which compensates for the factor of $g$
from quark-quark-gluon vertex.

Experimentally, it may be an easier task to look for a  
photon and a jet pair rather than a single jet. This was studied in the
context of heavy ion collisions in \cite{WangXN}. Therefore one can directly 
measure the correlator $C({\imb l}_\perp)$ that describes the
interactions of a high energy probe with the target nucleus. It should be 
noted that saturation effects become larger in the forward region
and therefore easier to measure experimentally. It is straightforward
to extend this calculation to the case of production of dileptons and
di-photons. This is currently under investigation.

Another interesting extension of this work would be to study the radiation of
a gluon by a quark as it moves through the classical field of the
nucleus. In that case, the emitted gluon can also interact with the
background classical field, which makes the calculation of this process
more involved.

\section*{Acknowledgment}
We would like to thank D. Bodeker, A. Dumitru, E. Iancu, L. McLerran,  
R. Pisarski, R. Venugopalan and W. Vogelsang for useful discussions.
F.G. is supported by CNRS.
J.J-M. is supported in part by a PDF from BSA and by U.S. Department 
of Energy under Contract No. DE-AC02-98CH10886.

\bibliographystyle{unsrt}
%\bibliography{biblio}

\end{document}